\documentclass[prl,twocolumn]{revtex4-1} 

\usepackage{graphicx}
\usepackage{amsmath}
\usepackage{amsfonts}
\usepackage{mathrsfs}
\usepackage{color}
\newcommand{\tn}{\textbf}

\def\<{\langle}
\def\>{\rangle}

\def\beq{\begin{equation}}
\def\eeq{\end{equation}}
\def\barray{\begin{eqnarray}}
\def\earray{\end{eqnarray}}

\begin{document}

\title{Entanglement of low-energy excitations in Conformal Field Theory}
\author{Francisco Castilho Alcaraz$^1$, Miguel Ib\'a\~nez Berganza$^2$, Germ\'an Sierra$^2$ \vspace{0.2cm} \\ $^1$ Instituto de F\'isica de S\~ao Carlos, Universidade de S\~ao Paulo, Caixa Postal 369, S\~ao Carlos, SP, Brazil.\\ $^2$ Instituto de F\'isica Te\'orica UAM/CSIC, Universidad Aut\'onoma de Madrid, Cantoblanco 28049, Madrid, Spain.}

\begin{abstract}
In a quantum critical chain, the scaling regime of the energy and
momentum of the ground state and low lying excitations are described
by conformal field theory (CFT). The same holds true for the von Neumann and
R\'enyi entropies of the ground state, which display a universal
logarithmic behaviour depending on the central charge. In this letter
we generalize this result to those excited states of the chain that
correspond to primary fields in CFT. It is shown
that the $n$-th R\'enyi entropy is related to a 2$n$-point correlator of
primary fields. We verify this statement for the critical $XX$ and
$XXZ$ chains. This result uncovers a new link between quantum
information theory and CFT.
\end{abstract}

\maketitle

Entanglement 
is one of the central concepts
in quantum physics since Schroedinger used the term in an answer to
the Einstein-Podolsky-Rosen article in 1935. A particularly active
line of research is concerned with the role played by entanglement
in the physics of many-body systems\cite{Amico}. One is
typically interested in the amount of entanglement between two spatial
partitions, say $A$ and $B$, of a many-body system in its ground
state. For a pure ground state the amount of entanglement is usually
quantified with the \textit{entanglement entropy}, or the von Neumann
entropy of the reduced density matrix $\rho_A$:
$S_A=-\mbox{tr$_A$ }\rho_A \ln \rho_A$. Alternatively, the R\'enyi
entropies $S_n$ are also used: $(S_n)_A=\frac{1}{1-n}\ln\mbox{tr$_A$
}{\rho_A}^n$, the entanglement entropy being $\lim_{n\to1} S_n$. One
of the most important results in this topic is the celebrated
\textit{area law}\cite{Srenidcki93,Eisert,Cirac09}, which, roughly
speaking, states that ground states of gapped many-body systems with
short-range interactions have an entanglement entropy proportional to the area of the hypersurface separating both
partitions
. The area law restricts the fraction of
the Hilbert space accessible to ground states of local Hamiltonians in
an essential way, allowing for their efficient numerical simulation \cite{Cirac09}.\\
\indent
Violations of the area law occur in gapless (critical) systems.  In
one dimension most of critical systems, as well as being gapless, are
also conformal invariant. The attention to the entanglement properties
on these systems came after the seminal result of Holzhey, Larsen and
Wilczek \cite{HLW}, who showed that the leading behavior of the ground
state entropies $S_n^{\rm gs}$ is proportional to the central charge of the
underlying conformal field theory (CFT) governing the long-distance
physics of the discrete quantum chain. If $\ell$ and $N$ are the
lengths of the partition $A$ and of the total system, both measured in
lattice spacing units, then the R\'enyi entropy of the ground state,
with periodic boundary conditions, is \cite{HLW,Calabrese04,Vidal2003}

\beq
S_n^{\rm gs}(\ell) = \frac{c\;(n+1)}{6n}  \ln \left[    \frac{N}{ \pi} \sin \left( \frac{ \pi \ell}{N} \right) \right]+\gamma_n
\label{HLWlaw}
\eeq
where $c$ is the central charge of the CFT and $\gamma_n$ is a non-universal constant. 
\\
\indent
In a critical model, the finite-size scaling of the energy of excitations is
given by the scaling dimension of the corresponding conformal operators
\cite{Cardy}. This fact suggests that also the entanglement entropy could be related
to properties of these operators. 
Entanglement of excited states has been considered previously. In \cite{Alcaraz08}  it was 
shown that the negativity of the excited states in the XXZ critical model 
shows a universal scaling. In \cite{Masanes}  it was shown that a violation of area 
law should be expected for the low lying excited states of critical quantum 
chains, and in \cite{Alba09},  it was considered the entanglement of very large energy 
excitations in XY and XXZ spin chains. 

\begin{figure}[h!]
\begin{center}
\begin{tabular}{cc}
\includegraphics[width= 4 cm]{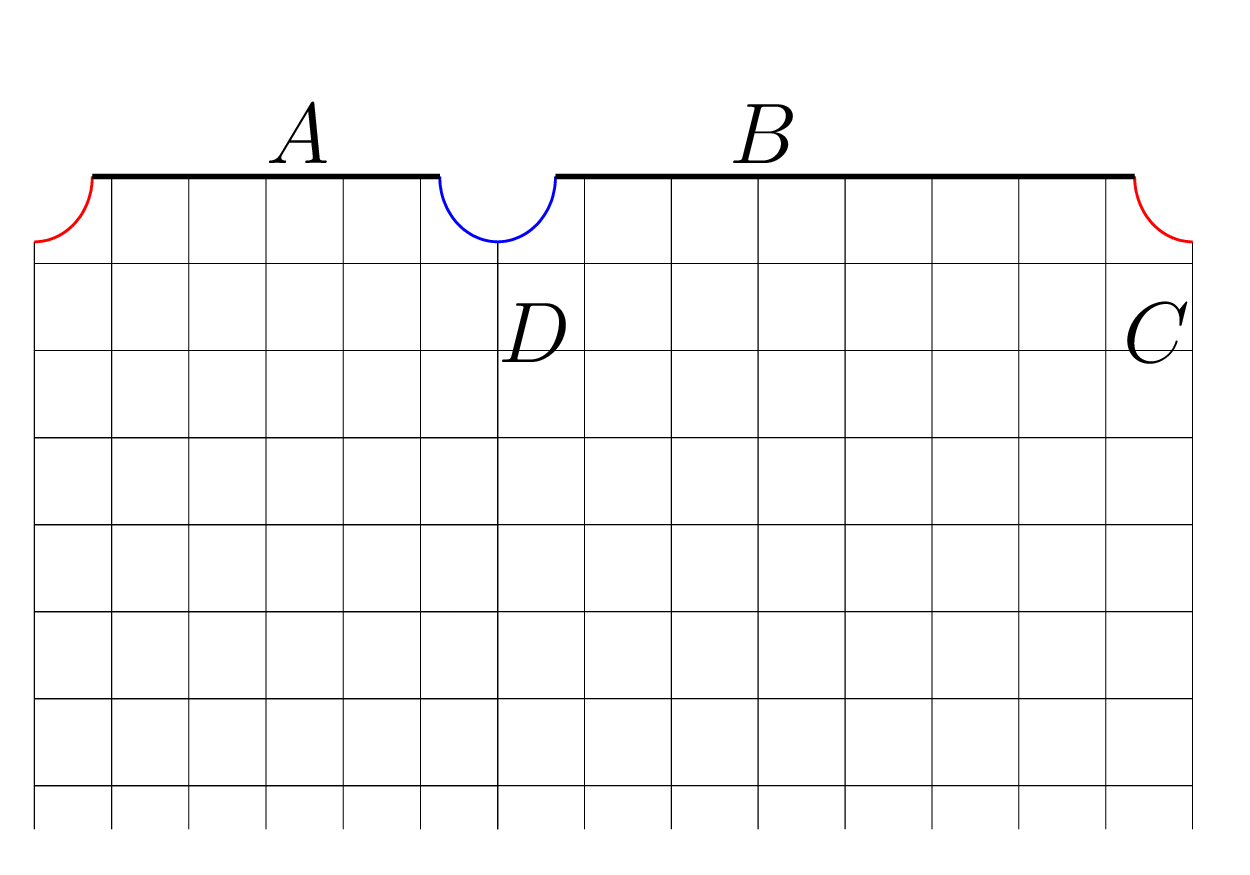} &
\includegraphics[width= 4 cm]{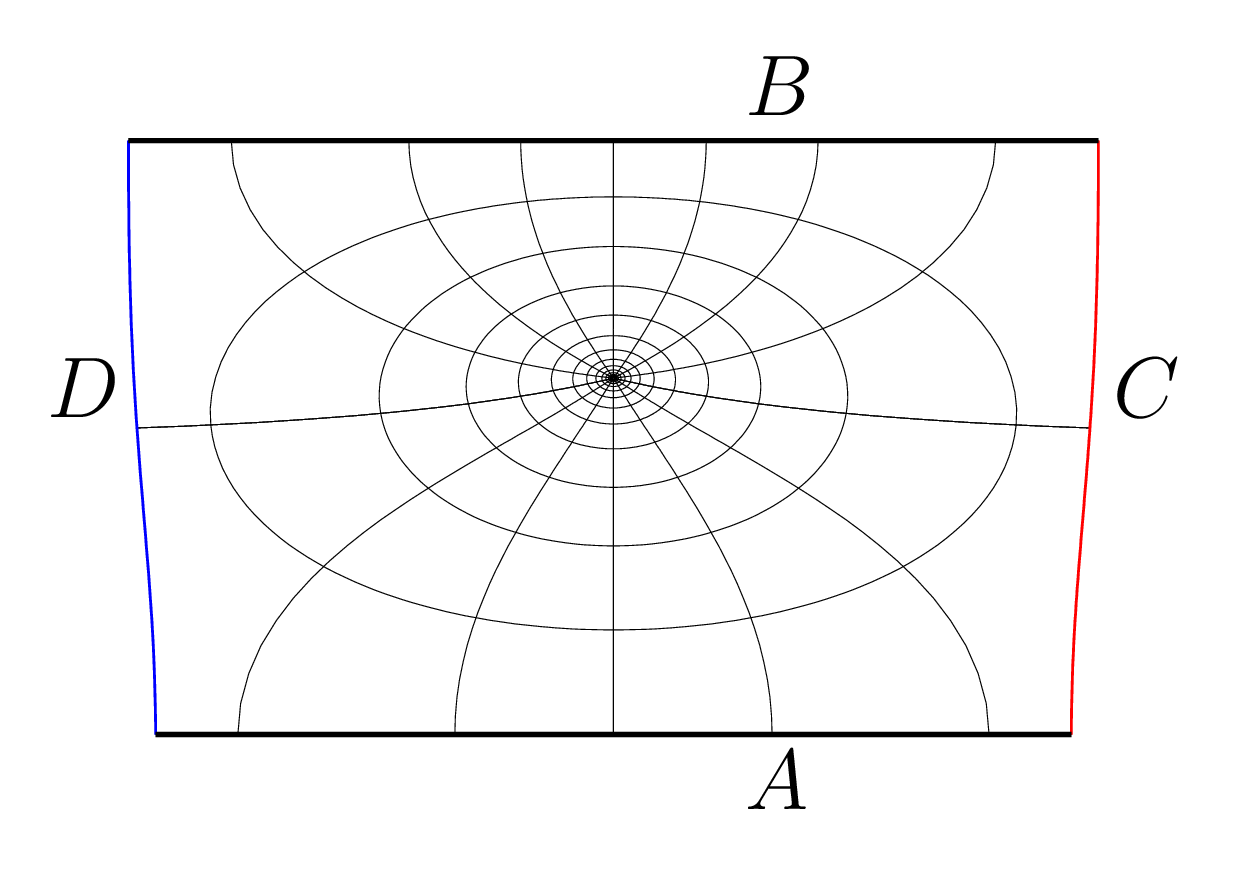} \\
$\zeta$ & $z$
\end{tabular}
\end{center}
\caption{Riemann surfaces describing the past events in $\zeta$ and $z$. The distinguished point in $z$ is the infinite past $\zeta_\infty= -i \infty$.}
\label{maps}
\end{figure}

In this letter we show that the entropy $S_n^{\rm exc}$ of excited
states associated to primary fields exhibits a universal behaviour
that generalizes (\ref{HLWlaw}). The energy of these low-lying states degenerate as
$1/N$ in the bulk limit $N\to\infty$. We prove that the
excess of entanglement, $S_n^{\rm exc}-S_n^{\rm gs}$, is a finite-size
scaling function related to the $2n$-point correlator of the primary field. These results are verified in two models: the $XX$ and $XXZ$ spin chains.\\ \indent
\tn{Entanglement of generic primary states.} Let us consider a system ${\cal S}$ of length $N$ with
periodic boundary conditions. To describe it, we introduce the complex
variable $\zeta = \sigma + i t$, where $0 \leq \sigma \leq N $ is the spatial
coordinate and $t$ is the time coordinate. ${\cal S}$ is split into
two subsystems $ {\cal S} = A \cup B$, with $A = (\epsilon, \ell- \epsilon)$ and $B = (\ell+ \epsilon, N- \epsilon)$, and where $\epsilon << \ell < N$ is a short-distance cutoff \cite{HLW}. The world sheet of the past ($t < 0$), is a cylinder with
two semidisks of radius $\epsilon$ cut out (denoted $C$ and $D$ in figure \ref{maps}). The boundary of the world sheet
of figure \ref{maps} is given by the union $A \cup C \cup B
\cup D$. After the conformal transformations $\zeta \rightarrow w \rightarrow z$:
 
\beq
w = - \frac{  \sin \left( \frac{ \pi (\zeta - \ell )}{ N} \right) }{ \sin \left( \frac{ \pi \zeta}{N} \right) }, \qquad
z =  \log \, w
 \label{cmap}
\eeq

\noindent
the $\zeta$ cylinder gets mapped into a strip of height $\pi$ and width $d = 2 \log  \left[  \frac{N}{ \pi \epsilon}\sin \left( \frac{ \pi \ell}{N} \right)   \right]$; being $A$, $B$, $C$, $D$ the boundaries of the strip in $z$ space (see figure \ref{maps}). Moreover, the point at the infinite past $\zeta_\infty = -i \infty$ gets mapped into 
$\zeta_\infty\to z_\infty = i \pi\left( 1 -  \ell/ N \right)$. 
%
%
We shall consider the simplest excited states in a CFT, namely the primary states, which are those generated acting on the vacuum $|0\>$ with a primary field $\Upsilon(\zeta, \bar{\zeta})$, with conformal weights $(h,\bar{h})$,

\beq
|\Upsilon \rangle = \lim_{\zeta, \bar{\zeta} \rightarrow -i\infty}  \Upsilon(\zeta, \bar{\zeta}) \, | 0 \rangle
\label{incoming}
.
\eeq
The wave function of this state $| \Upsilon \rangle$ is given by the path integral

\beq
\Psi_{XY}(\Upsilon) \propto \int {\cal D} \phi \;  \Upsilon[ \phi(z_\infty)]    \; e^{ - S(\phi)} 
\label{psixy} 
\eeq
where $\phi$ denotes the local field whose Euclidean action is
$S(\phi)$. The field $\Upsilon$ is a functional of $\phi$, that is evaluated
at the infinite past $z_\infty$ in equation (\ref{psixy}) (recall equation
(\ref{incoming})). $X$ and $Y$ denote the values of the
field $\phi$ in the subsystems $A$ and $B$ respectively. Periodic
boundary conditions are imposed on the $C$ and $D$ edges
\cite{HLW}. If $\Upsilon$ were not primary, then equation (\ref{psixy}) would include additional terms generated by the conformal transformations (\ref{cmap}).\\
\indent
The density matrix $\rho \equiv \rho_A$ of subsystem $A$ is obtained by tracing over the variables in $B$:

\beq
\rho_{X X' }(\Upsilon)  \propto \int {\cal D} Y \;  \Psi_{XY}(\Upsilon)  \, 
\Psi_{YX'}^*(\Upsilon) 
\label{rhoxy}
.
\eeq
Plugging (\ref{psixy}) into (\ref{rhoxy}) one finds

\beq
\rho_{X X' }(\Upsilon)  =   \frac{  \int {\cal D} \phi  \;  \Upsilon[ \phi(z_\infty)]  \,  \Upsilon^*[ \phi(z'_\infty)]  \; e^{ - S(\phi)} }{ Z(1)  \langle \Upsilon(z_\infty) \, \Upsilon^\dagger( z'_\infty) \rangle} 
  \label{rhoxx}
\eeq
where $z'_\infty = i \pi ( 1 +  \ell/ N )$ represents the point at the infinite future. The functional integral is over a strip of height $2 \pi$ and width $d$, with boundary conditions $\phi=X$ on the lower edge and $\phi=X'$ on the upper edge. The normalization  factor is determined by the condition ${\rm tr} \, \rho =1$, which implies that $Z(1)$
is the functional integral with no operator insertion and the top and bottom
edges of the strip being identified (i.e. a torus partition function), and $\langle \Upsilon \, \Upsilon^\dagger \rangle$ is the
two point correlator on this torus. To compute the entanglement entropy one first computes the trace of $\rho_\Upsilon^n$, which is given by 

\barray
\mbox{tr }\rho_\Upsilon^n  =   \frac{ Z(n)}{ Z(1)^n}  \frac{ 
 \prod_{k=0}^{n-1}\langle \Upsilon(z_\infty+2i\pi k) \, \Upsilon^\dagger( z'_\infty+2i\pi k) \rangle_{\tau_n}}{
 {\langle \Upsilon(z_\infty) \, \Upsilon^\dagger( z'_\infty) \rangle_{\tau_1}}^n} \nonumber  \label{tra} \\
\earray
where $Z(n)$ denotes the partition function on a torus of lengths $2 \pi n$ and $d$, so that the moduli parameter is given by $\tau_n = 2 \pi i n/  d$, and where $\<\ldots\>_{\tau_n}$ denotes the expectation value in the $\tau_n$-torus.  Notice that the $2 n$-point correlator of fields  $\Upsilon , \, \Upsilon^\dagger$ depends on the ratio $\ell/N$ and on the moduli parameter.\\
To further proceed one uses the expression of the partition function $Z(n)$ of a general CFT with central charge $c$ for chiral and antichiral sectors of the theory, $Z(n) = Z(\tau, \bar{\tau}) = \mbox{tr }q^{ L_0 - \frac{c}{24}} \; \bar{q}^{ \bar{L}_0- \frac{c}{24}}$, with the nome $q = \bar{q} = \exp{( 2 \pi i \tau)}$.  In the limit $d>>1$, it is convenient to perform the modular transformation $\tau \rightarrow - 1/\tau$. The partition function is modular invariant and can be  easily evaluated in terms of the nome $\tilde{q} = e^{- 2 \pi i/\tau} = e^{ - \ d/n}$. In particular, for the ground state ($\Upsilon_0  = {\bf 1}$) one gets (up to a model-dependent factor $c_n=e^{(1-n)\gamma_n}$):

\beq
{\rm tr } \; \rho_{\Upsilon_0}^n  = 
 \frac{ Z(n)}{ Z(1)^n} \sim e^{ \frac{ c}{12} ( \frac{1}{n} - n) \,d } =
 \left[    \frac{N}{ \pi \epsilon}
\sin \left( \frac{ \pi \ell}{N} \right) \right]^{ \frac{c}{6} ( \frac{1}{n} - n) }
\eeq
as anticipated in (\ref{HLWlaw}). In the general case, equation
(\ref{tra}) depends on a $2n$-point correlator of the fields $ \Upsilon$ and
$\Upsilon^\dagger$ on a cylinder of length $2\pi n$ along the time direction. It is now convenient to
rescale this length to $2 \pi$. Afterwards, we shift the coordinates $z_j\to
z_j-i\pi(1-x)/n$ where $x=\ell/N$. Finally, we exchange $\sigma$
and $t$ coordinates in such a way that $z_j=2\pi j/n$ for $\Upsilon$
and $z_j=2\pi (j+x)/n$ for $\Upsilon^\dag$. The ratio between the excited and the ground state traces, $F^{(n)}_\Upsilon(x)=\mbox{tr } \rho_{\Upsilon}^n/\mbox{tr } \rho_{\Upsilon_0}^n$, becomes, from (\ref{tra}):

\barray
F^{(n)}_\Upsilon(x)    \equiv
 \frac{ n^{-  2 n ( h + \bar{h})}\langle  \prod_{j=0}^{n-1}  \Upsilon(\frac{2 \pi j}{n}) \, \Upsilon^\dagger( \frac{2 \pi (j + x)}{n})
 \rangle_{\rm cy}}{  
  \langle \Upsilon(0) \, \Upsilon^\dagger( 2 \pi x) \rangle_{\rm cy}^n} \nonumber   \label{tra3} \\
\earray
where $\<\ldots\>_{\rm cy}$ denotes the expectation value in a cylinder of length $2\pi$. 
Note that $F^{(n)}_\Upsilon=\exp\left[(1-n)(S_n^\Upsilon-S_n^{\rm gs})\right]$. 
The dependence of the entropies of the excited states on the $n$-point 
correlation functions was also observed in the ground state entropies of two 
disjoint segments of the quantum critical chains \cite{cala-et-al}. 
The entanglement entropy for the excited state $|\Upsilon\>$ can then be computed using the replica trick:

\beq
S_1^{\rm exc} = S_1^{\rm gs}  - \frac{ \partial  F^{(n)}_\Upsilon}{\partial n} |_{n=1}
 \label{SAU}
.
\eeq
In the limit $x<<1$, the terms $\Upsilon\Upsilon^\dag$ appearing in (\ref{tra3}) can be approximated by the operator product expansion (OPE) $\Upsilon \times \Upsilon^\dag={\bf 1}+\Psi+\ldots$ , finding:

\beq
 F^{(n)}_\Upsilon(x)\sim 1+\frac{h+\bar h}{3}\left(\frac{1}{n}-n\right)(\pi x)^2+O(x^{2\Delta_\Psi})
\label{lowxF}
\eeq
where $\Psi$ is the operator with the smallest scaling dimension,
$\Delta_\Psi$. The term of order $x^{2\Delta_\Psi}$ depends on the OPE
constants $C_{\Upsilon\Upsilon^\dag}^\Psi$ and on the expectation values
$\<\Psi(0)\Psi(\frac{2\pi j}{n})\>_{\rm cyl}$. If $\Delta_\Psi=1$, this
term is $O(x^2)$ as the first one in equation (\ref{lowxF}), and
eventually they may cancel one another, as we shall see in an example
below. If $\Delta_\Psi\ne 1$ one could use (\ref{lowxF}) to infer the quantities
$h+\bar h$, $\Delta_\Psi$ and $C_{\Upsilon\Upsilon^\dag}^\Psi$ from the numerical computation of the entanglement.\\
\indent Using equation (\ref{SAU}) one
finds, for the low-$x$ behaviour of the entanglement entropy
($\ell/N<<1$):

\beq
S_1^{\Upsilon}(\ell)- S_1^{\rm gs}(\ell)\sim\frac{2 \pi^2}{3}(h+\bar h) \left(\frac{\ell}{N}\right)^2+O\left(\frac{\ell}{N}\right)^{2\Delta_\Psi}
\label{lowxS}
.
\eeq
\\
\indent
Equations (\ref{tra3}-\ref{lowxS}) are the main results of
this letter. They relate the von Neumann and $n$-R\'enyi entropy of the excitation
represented by the primary operator $\Upsilon$ to the $2n$-point
correlators of $\Upsilon$ and $\Upsilon^\dag$ in the cylinder. Notice that the ratio
$F_\Upsilon^{(n)}$ does not depend on the non-universal constant $\gamma_n$, which is therefore common to $S_n^{\rm gs}$ and $S_n^{\rm exc}$.\\ 
\indent As an example of the laws (\ref{tra3},\ref{SAU}) we shall
consider a $c=1$ CFT given by a massless boson compactified on a
circle. The primary fields are given by the vertex operators
$\Upsilon_1[n,m]=e^{i (\alpha_+\phi+\alpha_-\bar\phi)}$ (being
$\phi$,$\bar \phi$ chiral and antichiral boson fields) where
$\alpha_{\pm}=n/2\sqrt \kappa\pm m\sqrt \kappa$, $\kappa$ is the compactification ratio, and $n,m\in \mathbb Z$. The scaling dimensions of these operators are $(\alpha_+^2+\alpha_-^2)/2=n^2/4\kappa+m^2\kappa$. Using the chiral correlator of vertex operators on the cylinder, $\<\prod_je^{i\alpha_j \phi(z_j)}\>_{\rm cy}=\prod_{j>k}\left[2\sin(z_{jk}/2)\right]^{-\alpha_j\alpha_k}$,
it turns out that

\beq
F_{\Upsilon_1[j,k]}^{(n)}(x)=1,\qquad \forall x,j,k.
\label{vertexF}
\eeq
Hence, all the excitations represented by vertex operators
have the same entropy as the ground state. This result is not in contradiction with (\ref{lowxF}) because, in this case, $\Delta_\Psi=\Delta_{\partial\phi}=1$ and both $O(x^2)$ terms in (\ref{lowxF}) cancel out due to the properties of the OPE constants. In fact, the cancellation happens in all order of $x$.\\
\indent
Next, we study the operator
$\Upsilon_2=i\partial\phi$. Using its correlator on the cylinder
$\<\partial\phi(z_1)\partial\phi(z_2)\>_{\rm cy}=
-\left[2\sin(z_{12}/2)\right]^{-2}$ and the Wick theorem, we get (in
terms of $s(x)\equiv \sin(\pi x/2)$):

\beq
F^{(2)}_{\Upsilon_2} (x)  
 =  1 - 2 s^2(x)+ 3 s^4(x) - 2 s^6(x) +s^8(x)
\label{3rdF2}
\eeq
and a more lengthy expression for $F_{\Upsilon_2}^{(n>2)}$. In the
low-$\ell/N$ limit, one finds that $F^{(n)}_{\Upsilon_2} (x) \sim 1 + ( \pi x)^2\left(1/n - n \right)/3$, which leads to an excess of entanglement entropy given by (\ref{lowxS}) with $(h,\bar h)=(1,0)$.

Realizations of both types of excitations in particular models will be now shown, and their amount of entanglement compared with the CFT predictions (\ref{vertexF},\ref{3rdF2}). \\
\tn{Excitations in the $XX$ and $XXZ$ models.} The Hamiltonian of the spin-1/2 $XXZ$ model is given by

\beq
H_{\mbox{\scriptsize xxz}}=-\frac{1}{2}\sum_{j=1}^N\left( \sigma_j^x\sigma_{j+1}^x + \sigma_j^y\sigma_{j+1}^y+\Delta \sigma_j^z\sigma_{j+1}^z \right),
\label {xxz}
\eeq 
where $N$ is even and periodic boundary conditions are assumed (for
$\Delta=0$ we get the $XX$ model). This model is integrable
\cite{Lieb1961} and gapless for $-1\leq \Delta <1$. The corresponding
CFT is given by the aforementioned bosonic CFT with
$\kappa=\frac{\pi}{2}\left[\pi-\cos^{-1}(-\Delta)\right]^{-1}$. The $XX$ model in the sector
with magnetization $M=\frac{1}{2} \sum_j\sigma_l^z$ can be mapped, through a
Jordan-Wigner transformation, into a system with $n_F=M+N/2$ free
fermions in a lattice of $N$ sites. We computed the entanglement and
R\'enyi entropies of several types of excitations in these models. This
task was achieved using the methods of references
\cite{Peschel2003,Vidal2003} in the free fermion problem and through
numerical exact diagonalization in the $XXZ$ case. \\
\indent
Let us first consider the vertex operator
$\Upsilon_1[0,m]$. In the free fermion model, the result (\ref{vertexF}) is
exact and can be proved analytically. Indeed, $|\Upsilon_1[0,m]\>$ corresponds
to the umklapp excitation
$\prod_{j=1}^md^\dag_{k_F+(2j-1)\pi/N}d_{-k_F+(2j-1)\pi/N}|0\>$, where $k_F=\pi n_F/N$ is the Fermi momentum, and where $|0\>$ is the Fermi state and
$d^\dag_k$ the fermionic creation operator with momentum
$k$. This state can be obtained from the Fermi state shifting all the momenta as $k\to k+2m\pi/N$. Such a shift produces a global phase factor in
the wavefunction in real space and, consequently, the entropy remains
unchanged. In the $XXZ$ model, the state $|\Upsilon_1[0,1]\>$ corresponds to the
ground state in the sector with $n_F$ spins up and total
momentum $P=2\pi n_F/N$. We observe that the prediction
$F^{(n)}_{\Upsilon_1}(x)=1$ holds, up to the oscillations expected for $n\ge 2$ \cite{Calabrese2010}, which in this case are of the order of $10^{-3}$ for systems with $N=30$ spins.\\ 
\begin{figure}[h]
\begin{center}
\includegraphics[height=5.5cm]{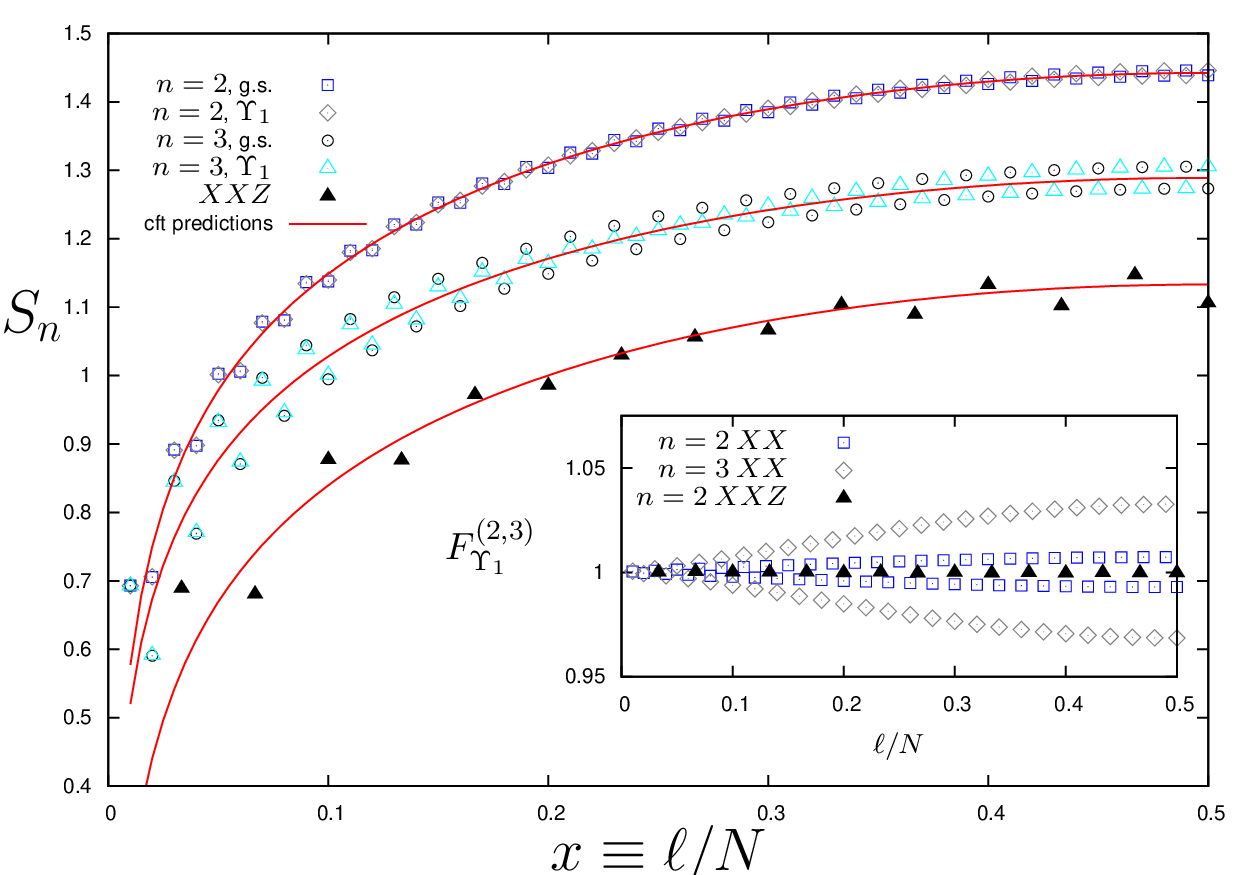}
\caption{{\footnotesize An illustration of the law $F_{\Upsilon_1[2,1]}^{(2)}(x)=F_{\Upsilon_1[2,1]}^{(3)}(x)=1$ for the $XX$ model with $N=100$, $n_F=50$, and for the $XXZ$ with $\Delta=-1/2$ and $N=30$, $n_F=14$ (16 for the excited state). The entropy of ground and excited states coincide and follows the law (\ref{HLWlaw}) (continuous lines), up to oscillations \cite{Calabrese2010}.}}
\label{1stFermions}
\end{center}
\end{figure}
We will now consider the excitation $\Upsilon_1[2,1]$. In a
system of free fermions the resulting state corresponds to
the addition of two fermions at the right of the Fermi point, i.e., to
the state $d_{k_F+\pi/N}^\dag d_{k_F+3\pi/N}^\dag|0\>$. Figure \ref{1stFermions} shows that ground and excited states
entropies $S_{2,3}$ coincide, up to oscillations. In the $XXZ$ model, $|\Upsilon_1[2,1]\>$ is the lowest eigenstate with total momentum $P=2\pi(n_F+2)/N$. Again in this case, oscillations of $F^{(2)}_{\Upsilon_1[2,1]}$ around one are observed (see figure \ref{1stFermions}).\\
\begin{figure}
\begin{center}
\includegraphics[height=5.5cm]{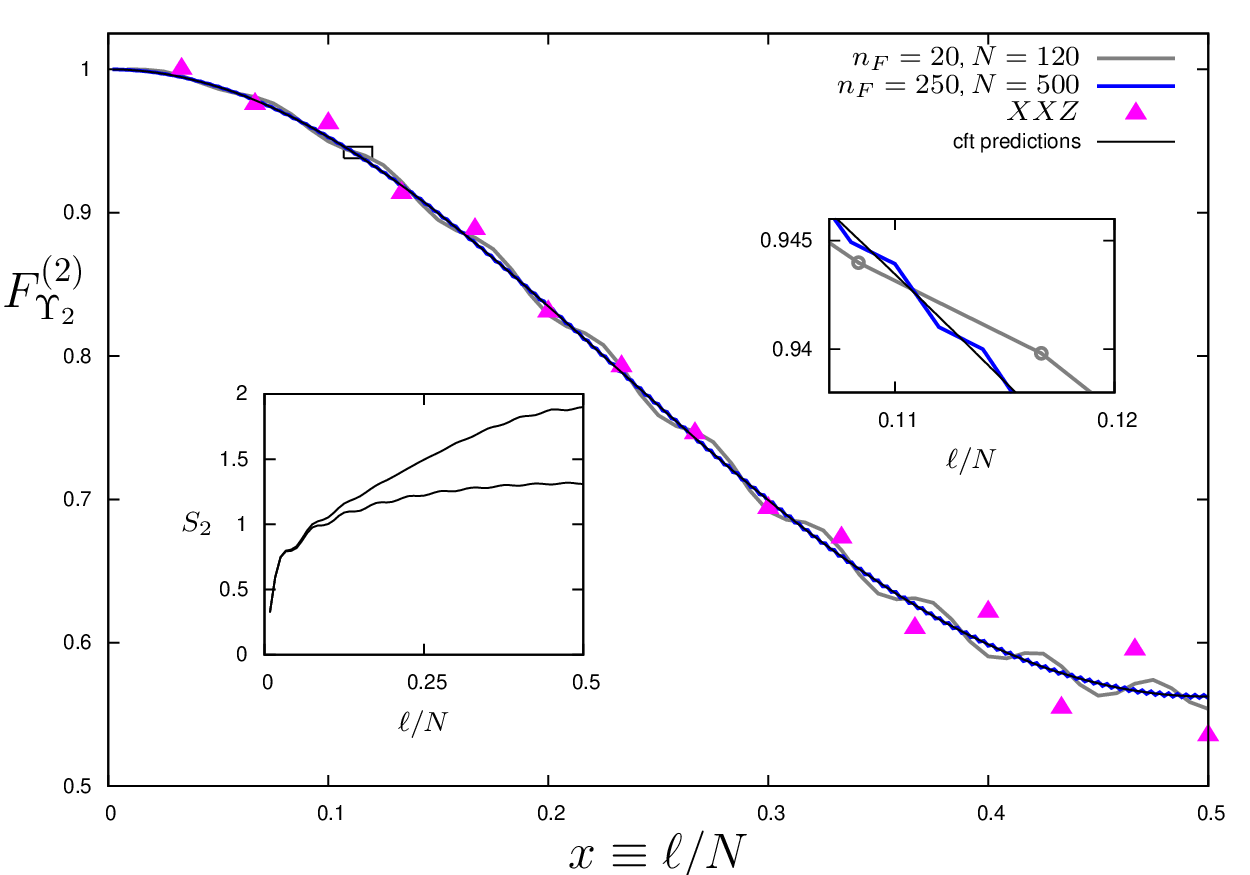}
\caption{{\footnotesize The quantity $F_{\Upsilon_2}^{(2)}$ for the $XX$ model at different filling fractions and for the $XXZ$ ($\Delta=-1/2$, $N=30$, $n_F=14$) model, vs. the CFT prediction (\ref{3rdF2}). For $n_F=250$ the oscillations around (\ref{3rdF2}) are so small that both curves are indistinguishable. In the inset we show $S_2$ for the ground and excited states ($n_F=250$, $N=500$). The upper inset is a zoom of the region selected by the small rectangle over the curve in the main figure.}}
\label{3rd}
\end{center}
\end{figure}
\indent
Finally, figure \ref{3rd} shows some numerical results for the entanglement of the excitation $\Upsilon_2=i\partial\phi$. In the free fermion problem, $|\Upsilon_2\>$ corresponds to a particle-hole excitation: $d^\dag_{k_F+\pi/N}d_{k_F-\pi/N}|0\>$, while in the $XXZ$ model it corresponds to the lowest eigenstate with $P=2\pi/N$. We observe an excellent agreement with the theoretical prediction (\ref{3rdF2}) for $n=2$. Similar results hold for $n=3$. Moreover, we have checked, for $n$ up to 6, that the low-$\ell/N$ formula (\ref{lowxS}) is very well satisfied for fermions.
\indent
In summary, we have obtained an expression for the R\'enyi entropies of excitations associated to any primary field. We verified the results with finite-size realizations of the $XX$ and $XXZ$ models up to 30 sites in the latter case, finding very good agreement with the theory. \\
\indent
As explained earlier, equation (\ref{tra3}) can be used as a numerical method to extract information
about correlators, conformal dimensions and OPE coefficients of primary fields. An interesting problem is to
generalize these results to the descendent states in CFT. We expect
that the R\'enyi entropies, at a given level of a conformal tower will
depend on the particular state targeted. This can provide a method to
establish the correspondence between degenerated excited
states of a critical lattice model, and the descendent fields in the
underlying CFT.\\
\indent
Equation (\ref{tra3}) further suggests a generalization of the R\'enyi entropies in terms of traces of different density matrices
$\mbox{tr}\left[\rho_{\Upsilon_1}\rho_{\Upsilon_2}\ldots\right]$. This
object would be related to the correlator:
$\<\Upsilon_1\Upsilon_1^\dag\Upsilon_2\Upsilon_2^\dag\ldots\>$
in the very same fashion as in (\ref{tra3}). The numerical computation of
the associated \textit{generalized entropies} would then provide
information on more general correlators in CFT, and vice-versa. Applications of the present work to other models and to non-primary fields are in progress. \\
\indent
This work represents a further step along the direction of deriving CFT data using quantum information methods.\\
\indent \tn{Acknowledgements.} 
We thank P. Calabrese, R. Pereira and V. Rittenberg  for discussions. This work was  supported by the Spanish project FIS2009-11654 and by FAPESP and CNPq (Brazilian agencies).


\begin{thebibliography}{3}
\bibitem{Amico} L. Amico, R. Fazio, A. Osterloh, V. Vedral, Rev. Mod. Phys. \tn{80} 517–576 (2008).
\bibitem{Srenidcki93} M. Srenidcki, Phys. Rev. Lett \tn{71} 666 (1993).
\bibitem{Eisert} J. Eisert, M. Cramer, M.B. Plenio, Rev. Mod. Phys. \tn{82} 277 (2010).
\bibitem{Cirac09} J. I. Cirac and F. Verstraete 2009 J. Phys. A: Math. Theor. \tn{42} 504004 (2009).
\bibitem{HLW} C. Holzhey, F. Larsen and F. Wilczek, Nucl Phys. B \tn{424} 443 (1994).
\bibitem{Calabrese04} P. Calabrese, J. Cardy, J. Stat. Mech. P06002 (2004).
\bibitem{Vidal2003} G. Vidal, J. I. Latorre, E. Rico, A. Kitaev, Phys. Rev. Lett. \tn{90} 227902 (2003).
\bibitem{Cardy} H. W. J. Bl\"ote, J. L. Cardy, M. M. Nightingale, Phys. Rev. Lett. \tn{56} 742 (1986); I. Affleck, Phys. Rev. Lett. \tn{56} 746 (1986).
\bibitem{Alcaraz08} F. C. Alcaraz, M. S. Sarandy, Phys. Rev. A \tn{78} 032319 (2008).
\bibitem{Masanes} L. Masanes, Phys. Rev. A {\bf 80}  052104  (2009).
\bibitem{Alba09} V. Alba, M. Fagotti, P. Calabrese, J.Stat.Mech. 10020 (2009).
\bibitem{cala-et-al} P. Calabrese, J. Cardy, and E. Tonni, J. Stat. Mech P011001 (2009); J. 
Stat. Mech. P01021 (2011); M. Headrick, Phys. Rev. D {\bf 82} 126010 (2010).
\bibitem{Peschel2003} I. Peschel, J. Phys. A: Math. Gen. \tn{36} L205 (2003).
\bibitem{Calabrese2010} P .Calabrese, M. Campostrini, F. Essler, B. Nienhuis, Phys. Rev. Lett. 104, 095701 (2010); 
 P. Calabrese and F. H. L Essler,  J. Stat. Mech. P08029 (2010).
\bibitem{Lieb1961} E. H. Lieb, T. Shultz, D. J. Mattis, Ann. Phys. \tn{16} 407 (1961).
\end{thebibliography}
\end{document}